\newcommand{\ket}[1]{\ensuremath{\left|  #1 \right\rangle}}

\documentclass[aip,rsi,reprint]{revtex4-1}

\usepackage{graphicx}
\usepackage{epstopdf}
\usepackage{hyperref}
\usepackage{fancyhdr}

\begin{document}

\title{A low phase noise microwave source for atomic spin squeezing experiments in $^{87}$Rb}

\author{Zilong Chen}
\author{Justin G. Bohnet}
\author{Joshua M. Weiner}
\author{James K. Thompson}
\affiliation{JILA, National Institute of Standards and Technology, and Department of Physics, University of Colorado, Boulder, Colorado 80309-0440, USA}
\date{\today}

\begin{abstract}
We describe and characterize a simple, low cost, low phase noise microwave source that operates near 6.800~GHz for agile, coherent manipulation of ensembles of $^{87}$Rb.  Low phase noise is achieved by directly multiplying a low phase noise 100~MHz crystal to 6.8~GHz using a non-linear transmission line and filtering the output with custom band-pass filters.  The fixed frequency signal is single sideband modulated with a direct digital synthesis frequency source to provide the desired phase, amplitude, and frequency control. Before modulation, the source has a single sideband phase noise near -140~dBc/Hz in the range of 10~kHz to 1~MHz offset from the carrier frequency and -130~dBc/Hz after modulation. The resulting source is estimated to contribute added spin-noise variance 16~dB below the quantum projection noise level during quantum nondemolition measurements of the clock transition in an ensemble $7\times 10^5$~$^{87}$Rb atoms.\\
\href{http://dx.doi.org/10.1063/1.3700247}{http://dx.doi.org/10.1063/1.3700247}
\end{abstract}


\maketitle

\section{Introduction}
Low phase noise microwave sources are crucial for a broad range of applications, including radar, communications, navigation, and timing-keeping~\cite{Santarelli1998}. For quantum sensors~\cite{Kasevich91,Canuel06}, quantum gates~\cite{Isenhower2010,Ospelkaus2011}, and tests of fundamental physics~\cite{Fortier07,Ashby07}, microwaves are often applied to perform rotations on pseudo-spins formed from two atomic hyperfine energy levels separated by spin-flip resonant frequencies in the microwave regime. 

In many atomic physics applications, the microwave source can be narrowband in its tuning range, since it is being used to address extremely stable atomic transitions.  However, the source must have nimble phase control in order to implement composite pulse sequences~\cite{TCS85, Wimperis94, CLJ03} that are used to cancel slowly varying amplitude and detuning errors.   When resonant microwaves are applied at the spin-flip frequency,  one can view the pseudo-spin as being rotated about a fixed axis in the x-y plane, with the rotation axis's azimuthal angle set by the phase of the microwaves. Composite pulses reduce the impact of amplitude and/or detuning errors for a single rotation about an axis by implementing several rotations about several axes with different azimuthal angles.  Thus, it is necessary to be able to quickly change the microwave phase between several values.  Conversely, if the microwave phase randomly fluctuates during the rotation, the orientation of the pseudo-spin at the end of the rotation may fluctuate from one realization to the next, swamping quantum sources of orientation noise, such as quantum projection noise, that are of fundamental interest.~\cite{WBI92,WBI94,FKJ08,AWO09,SLV10,LSV10,CBS11} 

Here we present the details of a custom, low phase noise microwave source near 6.8 GHz, with tuning bandwidth of approximately 40 MHz, where rapid changes of phase, frequency, and amplitude are possible.  The 40 MHz tuning range is sufficient to span hyperfine transitions in the widely used $^{87}$Rb atom, and only minor modifications are needed for operation with other commonly used alkali-atoms such as $^{133}$Cs, $^{85}$Rb, etc.  In the range of 10 to 100 kHz offset frequencies from the carrier, the single sideband phase noise of the source is nearly 20~dB lower than a commonly used YIG oscillator phase locked to the same crystal reference. 

The offset frequency range 10 to 100~kHz is particularly relevant for recent spin-squeezing experiments~\cite{CBS11} based on quantum nondemolition measurements of the pseudo-spin projection $J_z$ onto $\hat{z}$ of the ground hyperfine clock states \ket{F=2, m_F=0} and \ket{F=1, m_F=0} of an ensemble of nearly $10^6$ $^{87}$Rb atoms.    In particular, during a crucial $\pi$-rotation one finds that only phase noise spectral components near the Rabi-frequency $f_R\approx 40$ kHz contribute significant noise into $J_z$~\cite{CBW12}, whose value must be resolved below the projection noise level.

Compared to other microwave frequency sources with the necessary stability and agility, the source we present is also a cost-efficient option.  The total component cost is below $\$3,300$, when the Rb atomic reference is excluded.  The source achieves lower phase noise than many broadband commercial microwave synthesizers, such as the Agilent E8257D~\footnote{Part numbers are given as technical information only, and do not represent endorsement by NIST.}, which is also roughly an order of magnitude more expensive.  Other narrow-band microwave oscillators, such as cryogenic sapphire microwave oscillators~\cite{CBL05}, achieve much lower phase noise, but again at much higher cost and limited availability.  Related low phase noise microwave sources based on a dielectric resonator oscillator (DRO) phase locked with a few 100~kHz bandwidth to a non-linear transmission line (NLTL) frequency comb have been reported at the $^{133}$Cs 9.2~GHz~\cite{BGC09, Rossetto2011} and the $^{87}$Rb 6.8~GHz~\cite{RLR09} hyperfine transition frequencies.

\section{Low Noise, Fixed Frequency Source}

\begin{figure*}
\includegraphics[width=6in]{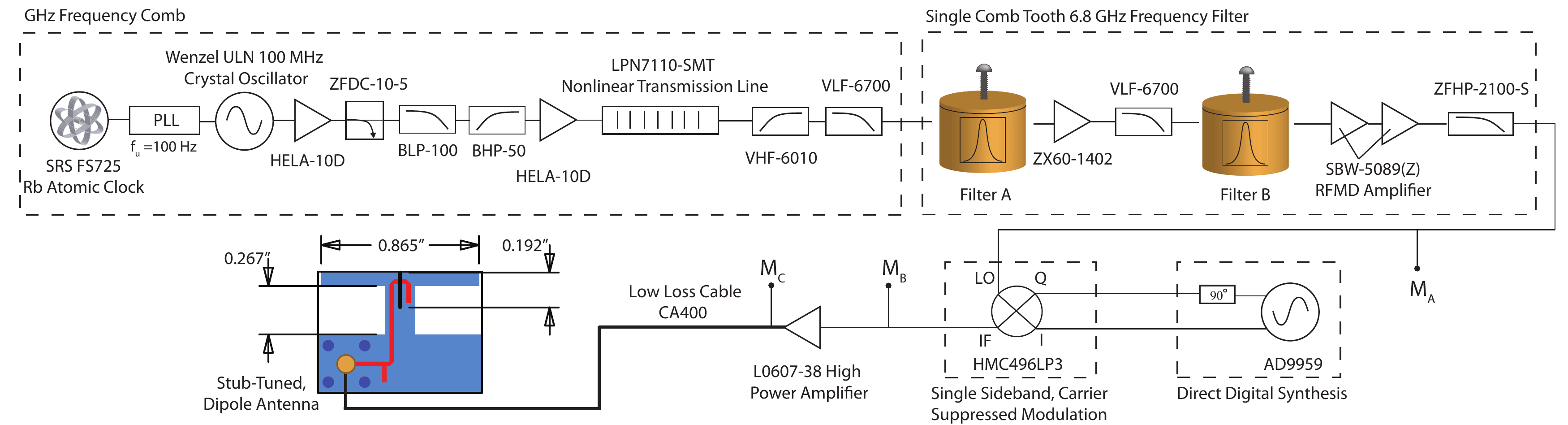}
\vspace{0.1in}
\caption{(color online) Schematic of low phase noise microwave synthesis chain. The source starts with a 100 MHz crystal oscillator phase locked to a 10 MHz atomic lock to provide very low frequency stability.  The 100 MHz signal, after coupling off a monitor output, is amplified and multiplied up through a NLTL.  The 68th harmonic is selected by a custom bandpass filter built from a series of commercial filters and homemade microwave resonators.  The resulting NLTL source is a single harmonic that provides the LO input for a single sideband modulator.  The I and Q inputs of the modulator are driven by a direct digital synthesis (DDS) board that allows phase coherent control over the frequency, amplitude, and phase of the microwave signal sent to the high power microwave amplifier.  The microwaves are coupled to the atoms using a printed circuit board dipole antenna. The signal line on the antenna is in red, and the ground plane, on the opposite side of the board, is in blue. The labels $\mathrm{M_A, M_B~ and~M_C}$ indicates points where phase noise was measured.} 
\label{fig:SourceFigure}
\end{figure*}

The microwaves are generated by first multiplying up a low phase noise crystal source to generate low noise microwaves at 6.800~GHz and then performing single sideband modulation of the carrier using a direct digital synthesis board as the modulation source (Fig.~\ref{fig:SourceFigure}).  A low noise 100~MHz crystal oscillator (Wenzel ULN P/N: 501-16843) is phase locked to a 10~MHz Rb clock signal (Standford Research Systems P/N: FS725) for long term frequency stability, with servo unity gain frequency of 100~Hz.   The 100~MHz signal is amplified to +27 dBm (Minicircuits P/N: HELA-10D) to drive a NLTL (Picosecond Pulse Labs P/N: LPN7110-SMT) generating a comb of harmonics out to 20~GHz.  For noiseless multiplication, the timing jitter is preserved, but the phase noise is increased relative to the source by $20\log_{10}(n)$~dB, where $n$ is the harmonic order.  We utilize the n=68 harmonic at 6.8~GHz yielding a fundamental phase noise increase of 36.7 dB over the crystal source.  Note that the frequency comb produced by the NLTL source allows for the flexibility to choose a frequency relevant for other alkali atoms by choosing a different harmonic to filter.

The harmonic at 6.8~GHz is isolated via a combination of standard filters (Minicircuits P/N: VHF-6010, VLF-6700, ZFHP-2100-S) for broadband filtering and two high quality factor ($Q$) microwave resonators~\cite{pipecap1} for narrow-band filtering. At 6.8~GHz, the measured insertion losses of the VHF-6010, VLF-6700, and ZFHP-2100-S  are 0.7, 0.8, and 0.3~dB respectively. The insertion losses are low enough that the thermal noise floor of -174~dBm/Hz at 300~K does not affect the phase noise of the synthesis chain. The ZFHP-2100-S, nominally a high pass filter with 3~dB point at 2.1~GHz, was used to attenuate high frequency harmonics above 14~GHz.  The VHF-6010 high pass 3~dB point is at 6.01~GHz, and the VLF-6700 low pass 3~dB point is at 7.6~GHz.  At an insertion loss of $\mathrm{IL} = 2.8(2)$ dB, the loaded quality factor is $Q_L \approx 580$, and the loaded half width at half maximum is $\approx 6$~MHz, much smaller than the 100~MHz spacing between the comb harmonics.  The microwave resonators have an unloaded $Q_0  = Q_L / \left(1 - 10^{-\mathrm{IL}/20} \right) \approx 2100$. Two microwave resonators are used to give strong suppression of the harmonics close to the harmonic at 6.8~GHz.

\begin{figure}
\vspace{0.1in}
\includegraphics[width=3.4in]{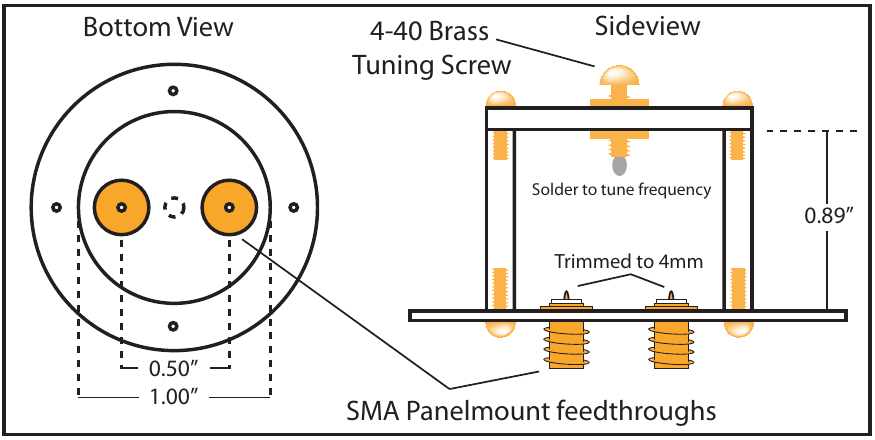}
\caption{(color online)  Diagram of the high-Q resonant microwave filters. The resonant cavity is constructed from a copper pipe, with copper plate used to form the lid and base.   The cavity resonance frequency is tuned with a 4-40 brass screw, secured tightly for good conductivity by a nut on either side of the center of the lid, and more finely tuned through a small amount of lead solder attached to the tip of the screw (see text). The signal is injected and extracted through panel mount SMA feedthroughs, trimmed to provide $2.8$~dB of insertion loss while maintaining a loaded $Q$ of about 580.}  
\label{fig:layout}
\end{figure}

The resonators (Fig.~\ref{fig:layout}) are constructed from stock copper pipe (inner diameter of $1"$, outer diameter of $1.5"$),  with a lid $0.25"$ thick made from copper barstock and a base from stock $0.125"$ copper sheet.  The three pieces are held together by screws that allow for a tight fit and good conductivity, but still provide the ability for the resonator to be disassembled.  We found no major improvements to the unloaded resonator $Q$ by soldering the pieces together.

The coupling of microwaves into and out of the resonator is provided by two panel mount SMA connector feedthroughs held to the copper sheet with a lock washer and nut.  Trimming the internal signal pin lengths increased the $Q$ at the expense of increasing the insertion loss.  For example, pins that had been trimmed to be $3$ mm above the base plate had an insertion loss of $5.2$ dB, but trimming down to $2$ mm above the base plate resulted in an insertion loss of $18.9$ dB.  The filters used for the final NLTL source were trimmed to be 4 mm above the base plate.

Coarse tuning of the resonant frequency was performed by changing the depth of a brass screw inserted in the center of the lid. We found that good conductivity between the screw and the lid was one of the most important parameters for obtaining the maximum $Q$.  The $Q$ was also reduced when using a stainless steel tuning screw in place of the brass screw.  After coarse tuning  the resonance frequency to within 20 MHz of the desired 6.800 GHz, two brass nuts were placed on either side of the screw to firmly secure it in place. The final fine-tuning of the center frequency was performed by attaching a small amount of Sn$_{0.6}$Pb$_{0.4}$ solder to the tip of the tuning screw, then successively sanding small portions of the solder away until the resonator reached the desired frequency.  Tuning resolution below 2~MHz was possible with this approach.

The center frequency of the two resonators were set to $6.800(2)$~GHz.  An amplifier (Minicircuits P/N: ZX60-1402) was placed between the resonators to prevent the formation of a coupled-resonance. After filtering, the largest unwanted comb tooth is at 6.7~GHz, $-49.5$~dB below the desired 6.8~GHz comb tooth.   The resonance frequencies can drift due to temperature changes with a coefficient of $\sim 150~\mathrm{kHz/K}$. Mechanical vibrations can modulate the resonance frequency, giving rise to a phase modulation of the transmitted microwaves to lowest order.  To mitigate both temperature drifts and vibration-induced phase modulation, we passively isolate the resonators in a foam box.  

The single sideband phase noise power spectrum of the 6.8~GHz NLTL source was measured by transporting the source to the National Institute of Standards and Technology (NIST) at Boulder, CO, and using a self-homodyne phase noise detector (OEwaves P/N: OE8000).  Fig.~\ref{fig:PhaseNoiseMeas}a shows the measured noise of our NLTL source at measurement point $\mathrm{M_A}$ in Fig.~\ref{fig:SourceFigure} (after filtering, but before modulation), a YIG oscillator (Micro Lambda Wireless P/N: MLPE-1290) phase locked to the same ULN crystal reference, and the measurement noise floor of the OE8000 calibrated using a Poseidon sapphire oscillator at 10~GHz. Our ability to measure phase noise is limited by the measurement noise floor below $\sim 500$~Hz offset frequencies,  thus the displayed data is an upper bound for the phase noise below $\sim 500$~Hz.

The NLTL source is quieter than the YIG source by $\sim 15$ to $20$~dB in the offset frequency range from 10~kHz to 400~kHz.  The multiplied-up phase noise of the 100~MHz ULN crystal taken from the manufacturer's datasheet agrees relatively well with the observed data, indicating that there is little added technical noise at this level in the multiplication, amplification, and filtering relative to the observed noise level. The 100~MHz ULN phase noise slope changes from $1/f^2$ above 1~kHz to $1/f^3$ below 1~kHz due to $1/f$ phase noise from the crystal oscillator feedback amplifier being converted into oscillator phase noise via the Leeson effect~\cite{Leeson1966}. The roll-off of the phase noise near 5~MHz is believed to arise from the high-$Q$ microwave filters.

\section{Agile Control of Phase, Frequency and Amplitude}

To generate amplitude, frequency and phase tunable microwaves at the hyperfine clock transition frequency 6.834693~GHz, a single sideband (SSB) modulator (Hittite P/N: HMC496LP3), with a manufacturer specified output noise floor of $-150$~dBm/Hz at 7 GHz, modulates the 6.8~GHz NLTL source.  The SSB modulator is tuned to give $-31$~dB and $-19$~dB suppression relative to the +1 sideband for the $0$, $-1$ and $-2$, $+2$ sidebands respectively.  

The I and Q modulation ports are driven near 34~MHz using $90^\circ$ relative phase signals from 2 of the 4 channels of a direct digital synthesis (DDS) frequency source (Analog Devices AD9959 evaluation board) operating near 34~MHz. The DDS source is phase stabilized to a 500 MHz reference provided by multiplying the 100~MHz ULN crystal oscillator using a diode pair (Avago P/N: HSMP-3822) and comb filtering provided by two bandpass filters (Minicircuits P/N: BPF-B503+). The DDS SSB phase noise floor specified by the manufacturer is $-151$~dBc/Hz at 10~kHz offset from the carrier when operated at 40~MHz with a 500~MHz reference signal, but this specified noise floor rises if the DDS is operated with a lower frequency reference, relying on a programmable internal PLL to multiply up the reference signal to 500~MHz. 

\begin{figure}
\vspace{0.1in}
\includegraphics[width=3.5in]{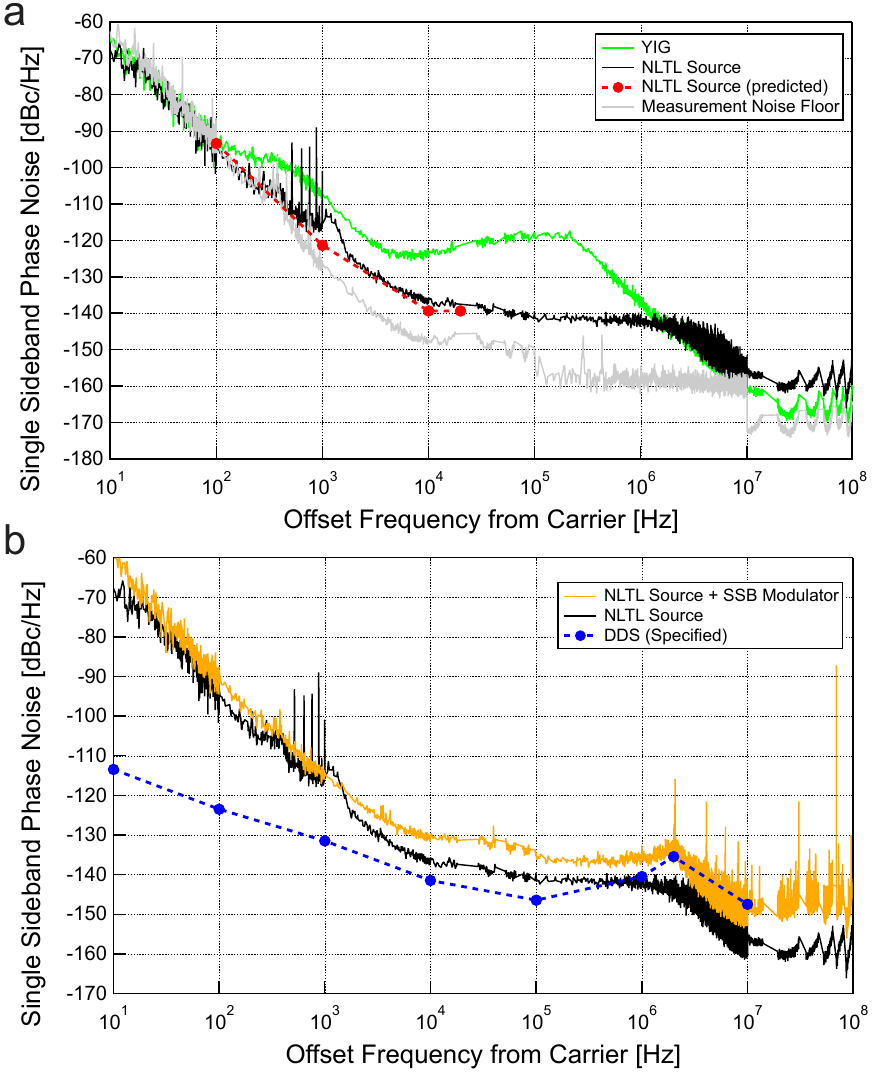}
\caption{(color online) Single sideband phase noise of several microwave sources all locked to or derived from the Wenzel ULN oscillator.  The measurement noise floor is in grey. {\bf a} Comparison of noise at 6.800~GHz for a YIG oscillator (green) and our microwave NLTL source (black) measured at point $\mathrm{M_A}$ in Fig.~\ref{fig:SourceFigure}.  The noise in the NLTL source agrees with the prediction (red circles with dashed line to guide the eye) based on ideal scaling of the noise in the ULN oscillator specified by the manufacturer. {\bf b} The modulated signal at 6.834~GHz (orange), measured at point $\mathrm{M_B}$ in Fig.~\ref{fig:SourceFigure}, exhibits higher phase noise than the 6.8~GHz signal from the NLTL source (black). The manufacturer specified noise of the DDS (blue circles with dashed line) when operated with an internal PLL of $\times 5$ is plotted to show that the added noise primarily comes from the SSB modulator, excluding frequencies above 2~MHz where the gain peak from the internal PLL dominates.  Spurious multiples of 60~Hz have been removed for visibility.} 
\label{fig:PhaseNoiseMeas}
\end{figure}

We control the DDS source through a custom LabVIEW interface that can update the phase, frequency and amplitude in 24, 32, and 28 $\mu$s respectively.  The tuning resolutions are 0.022$^\circ$, 0.12~Hz, and 0.1\% of full scale amplitude respectively.  Phase continuous frequency ramps and amplitude ramps are also possible which enable additional techniques, such as atomic population manipulation using Landau-Zener avoided crossings.  The LabVIEW interface utilizes 4-bit serial communication at 2~MHz, while the DDS is capable of communication at 125~MHz.  The communication rate is presently limited by the speed of the National Instruments DIO card (P/N: PCIe-6259) used to implement the serial communication.  If less flexibility or less integration with the data acquisition system is required, the DDS can also be programed through a manufacturer-provided USB interface and software, while still maintaining the ability to modulate a single parameter between as many as 16 programmable values. 

The SSB modulator output is amplified to as much as 8.9~Watts using a high power microwave amplifier (Microwave Power P/N: L0607-38). Measurements with the OE8000 at measurement point $\mathrm{M_C}$ in Fig.~\ref{fig:SourceFigure} show that the microwave amplifier adds a negligible amount of phase noise relative to the measured phase noise of the NLTL source derived from the ULN crystal shown in  Fig.~\ref{fig:PhaseNoiseMeas}a.  The microwave radiation is coupled to the atoms through a low loss $1.5$~m cable (CA400) and stub-tuned dipole antenna~\cite{antenna} fabricated directly on a high frequency laminate circuit board (Rogers P/N: RO3035).  The critical dimensions for the frequency tuning of the antenna are labeled in Fig.~\ref{fig:SourceFigure}. The antenna is relatively broadband, with a FWHM of 400~MHz. After careful alignment of the dipole antenna, we achieved a Rabi frequency of $f_R \approx 40$~kHz with the antenna approximately $1.5"$ from the atoms~\footnote{As a side note, microwave absorbing material (Laird RFLS Single Layer sheets, 0.25~inch thick) around the glass cell vacuum chamber was necessary to reduce  observed changes in the Rabi frequency of greater than 10\% peak to peak caused by people walking past the optics table.}.
 
The SSB modulator adds 5 to 8~dB of phase noise relative to the NLTL source, as seen in Fig.~\ref{fig:PhaseNoiseMeas}b where the measurement is taken at point $\mathrm{M_B}$ in Fig.~\ref{fig:SourceFigure}.  For frequencies below 2~MHz, we attribute the added noise to the SSB modulator.   This data was taken with the the DDS referenced to the 100~MHz Wenzel ULN crystal directly, and multiplication to 500~MHz being accomplished using the internal DDS PLL multiplier set to $\times 5$.  In this configuration, the phase noise of the DDS specified by the manufacturer, and confirmed by our own measurements, only has a small contribution at offset frequencies less than 2 MHz where an expected noise peak from the DDS PLL becomes visible.  This noise peak was present for the measurements in Fig.~\ref{fig:PhaseNoiseMeas}b,  but is not present for the final source that has a DDS reference clock of 500~MHz obtained via external $\times 5$ multiplication as described above.

\section{Impact on Quantum Nondemolition Measurements}

To achieve conditional spin squeezing using quantum nondemolition measurements of the atomic spin projection $J_z$ as described in ref.~17, it was necessary to apply a $\pi$-rotation using a microwave source.  Using the YIG PLL microwave source locked directly to the 10 MHz Rb atomic clock for the $\pi$-rotation, we observed added noise in the spin projection $J_z$ that was 18~dB above the quantum projection noise level for the $7 \times 10^5$ atoms.   From manufacturer specifications, the multiplied phase noise of the 10 MHz crystal is -95~dBc/Hz, for the relevant offset frequencies near $f_R$.

Utilizing the NLTL source locked to a slightly noisier $100$~MHz crystal oscillator (Wenzel Sprinter P/N: 501-04517D) with the SSB modulation (phase noise measurements taken but not shown), the predicted added noise noise in $J_z$ due to phase noise during the $\pi$-rotation is predicted to be $16$~dB below the projection noise level.  The prediction is made by simple scaling from the first measurement with the much noisier YIG oscillator.  By reducing the added noise from microwave rotations, we lowered the technical noise floor of the measurement and were able to observe a $5$~dB reduction in spin noise,  limited by our probe quantum backaction.   By changing the 100~MHz Sprinter oscillator to the ULN model and improving the added phase noise of the SSB modulator, future experiments could operate with added microwave phase noise to $J_z$ that is $27$~dB below the projection noise limit for this large ensemble size.  Lower microwave phase noise can be exploited to perform composite microwave pulses that would add more phase noise into $J_z$ than the simple $\pi$-rotation discussed here~\cite{CBW12}.

In conclusion, we have demonstrated a microwave frequency source with low phase noise that maintains the flexibility to adjust phase, frequency, and amplitude needed in high precision experiments with large atomic ensembles.  The source is very cost-effective, providing a phase stability that surpasses many commercial frequency sources for a fraction of the cost. Although the source is narrowband, the center frequency can be chosen for a particular need or atom that requires a stable microwave frequency, providing an attractive option for many applications.

\section{Acknowledgments}
We thank Archita Hati and David Howe for providing test equipment and help with the phase noise measurements. Z.~C. acknowledges support from A*STAR Singapore. J.~G.~B. acknowledges support from NSF GRF.  This work was supported by the NSF PFC and NIST.

%

\end{document}